\documentclass[journal,draftcls,onecolumn,12pt,twoside]{IEEEtranTCOM}
\usepackage{graphicx}
\graphicspath{{fig/}}
\usepackage{amsmath, amssymb, amsthm} 
\usepackage{graphicx}
\usepackage{cite}
\usepackage{subfigure}
\usepackage{mathtools} 
\usepackage{commath} 
\usepackage{multirow} 
\usepackage{enumerate} 
\usepackage{float} 
\usepackage{comment}
\usepackage{algorithm}
\usepackage{tabularx}
\usepackage{epstopdf}
\usepackage{empheq} 
\normalsize

\ifCLASSINFOpdf
\else
\fi

\begin{document}
%
\title{Multiple Access for Transmissions Over Independent Fading Channels}
%
%
%

\author{Bei-Hao Chang, Chia-Fu Chang, Pin-Wen Su, I-Hsien Yeh, Kai-Chuan Cheng, Ying-Chen Lin,  and Mao-Chao Lin~\IEEEmembership{Member,~IEEE,}
        
\thanks{ This work was financially supported by the Ministry of Science and Technology of Taiwan under Grants MOST 105-2622-8-002-002, and sponsored by MediaTek Inc., Hsin-chu, Taiwan}

\thanks{C.-F. Chang and I.-H Yeh are with MediaTek Inc., Hsin-chu, Taiwan, Taiwan. R.O.C. (e-mail: d93942016@ntu.edu.tw, lisatangchiu@gmail.com)}

\thanks{P.-W Su is with the School of Electrical and Computer Engineering at Purdue University, West Lafayette, IN 47907 (e-mail: su173@purdue.edu)}

\thanks{K.-C. Cheng, Y.-C Lin and B.-H Chang are with the Graduate Institute of Communication Engineering, National Taiwan University, Taipei 106, Taiwan, R.O.C. (e-mail: {r03942094,r04942118,r05942106}@ntu.edu.tw)}

\thanks{M.-C. Lin is with the Department of Electrical Engineering, National Taiwan University, Taipei 106, Taiwan, R.O.C. (e-mail: mclin@ntu.edu.tw)}        
}

%
%

\markboth{}%
{}
%



\maketitle

\begin{abstract}
We propose to employ a multilevel detection (MLDT) technique to allow multiple users which respectively transmit messages over independent fading channels to share the same resource, e.g., the same signature sequence in the CDMA (code division multiple access) system.   The users are separated by the different channel gains including amplitudes and phases resultant from the independent fading channels.  In a CDMA system with a fixed amount of available signature sequences, the number of users can be doubled or tripled by using MLDT although there is the cost of some BER performance degradation. 
\end{abstract}


%
\IEEEpeerreviewmaketitle

\section{Introduction}
%
%
%
%
\IEEEPARstart{M}{ultiple} access (MA) techniques are essential for multiple users to share the same bandwidth in the mobile communication systems, e.g., code division multiple access (CDMA) for the third generation (3G) mobile telecommunications and orthogonal frequency division multiple access (OFDMA) for the fourth generation (4G) systems. In recent years, many MA techniques have been proposed such as the Interleave division multiple access (IDMA) \cite{Li06}, non-orthogonal multiple access (NOMA) \cite{Saito13}, low-density signature CDMA (LDS-CDMA) \cite{Hoshyar08} and sparse code multiple access (SCMA) \cite{Nikopour13} and many more. Each of these MA works aims to either increase the sum rate of all users or serve more users in a limited bandwidth.

 For an MA system, each user is usually assigned with a unique resource so that the receiver can separate out the message of the desired user.  In the conventional CDMA system, the unique resource is the unique signature sequence.  In the IDMA system, the unique source is the unique interleaver.  In this paper, our goal is to find a technique to allow $P$ users to share the same resource, where $P \ge 1$.  In this paper, $P$ = 2 and 3 are studied.
Such a situation is referred to as user collision. In this way, the number of users in the CDMA system which is supplied with $K$ signature sequences can be multiplied to $KP$. This technique may also be applied to other MA systems for increasing the number of users.  

 The basic idea of our work is the observation that the channel gains of independent users are likely to be distinct in the uplink transmission of mobile communication system.  For NOMA in \cite{Saito13}, different users can be separated by distinct power levels.  In our work, not only the amplitude of the channel gain (equivalently, the power level) but also the phase of the channel gain will be utilized for separation of users. The superimposition of  signals from the $P$ collided users transmitted over independent fading channels can be viewed as a multi-level ($2^P$-level) signal. Exploiting the feature of the superimposed signal, the detector can recover the message for each of the collided users respectively.  Hence, such a technique is referred to here as the multi-level detection (MLDT) technique.  We specifically show the formulae of bit error rates (BERs) of uncoded MLDT for $P$ = 2 and 3.  The derived upper bounds indicate that at high SNR the BERs for $P$ = 2 and 3 are respectively 50\% and 133\% higher than that for $P$ = 1. 
 
 Error-correcting codes (ECC) are usually used to enhance the system reliability. For a $P$-user MLDT system in which the $P$ users employ $P$ identical binary LDPC (low density parity check) \cite{LDPC, Kschischang01} codes, the receiver can be implemented by using a MLDT device followed by $P$ binary decoders each of which uses a conventional sum product algorithm (SPA) to recover the message for a user.  Alternatively, the receiver can be implemented by using an MLDT device followed by a nonbinary generalized sum product algorithm (GSPA) \cite{PLNC} to recover messages of all the $P$ users.
The concept of MLDT is inspired by the work in \cite{PLNC}, of which the physical-layer network coding for the relay node can be considered as a coded  MLDT device for two users followed by a GSPA decoder.   

The asymptotic performances of coded MLDT can be evaluated by its capacity.  
Through the capacity performances of two-user MLDT using BPSK transmission over independent Rayleigh fading channels and a single-user using QPSK transmission over a Rayleigh fading channel, we see that the cost of two-user MLDT using BPSK transmission is the slightly reduced rate as compared to the single-user QPSK transmission.  In the single-path fading channel environment, using a fixed-rate ECC can achieve very little coding gain.  We will see that Raptor coded MLDT \cite{Shokrollahi} has the potential to achieve the coding gain implied by the capacity analysis.
 
The MLDT technique can be applied to CDMA systems and possibly other multiple access systems.  In this paper, we restrict our study to only CDMA systems.
In the CDMA system, orthogonal signature sequences, such as Hadamard-Walsh (HW) codes, can be employed to maintain the orthogonality among users. However, after the transmission over multipath fading channels, the sequences carried by different paths may no longer remain orthogonal and hence severe interference among users may occur. To alleviate the multipath interference, multi-carrier system can be considered. We apply the MLDT scheme to a multi-carrier direct-sequence CDMA (MC-DS-CDMA) system, where the spreading is performed for each subcarrier so that the orthogonality among users can be maintained and the number of users can be multiplied.

The remainder of this paper is organized as follows. Section \ref{sec:uncoded MLDT} introduces the proposed MLDT scheme and analyzes its BER performances.  In Section \ref{sec:coded MLDT}, capacity analysis together with some coded MLDT designs are provided. In Section \ref{sec:MLDT_CDMA}, the MLDT technique for CDMA systems is explored. 
We conclude the paper in Section \ref{sec:con}.

\underline{Notation}: Vectors are denoted by boldface case. $\text{Pr}\left\{\cdot\right\}$ and $\mathbb{E}\left[\cdot\right]$ stand for the probability and the expected value of a random variable, respectively.

\section{MLDT Receiver} \label{sec:uncoded MLDT}
Consider a symbol-synchronous communication system in which $P$ users share the same resource. For user $p$, let ${\bf b}_p = [ b_{p}(1),...,b_{p}(n),...b_{p}(N)]$ be the bit sequence to be transmitted, where $b_{p}(n) \in \{0, 1\}$.  Denote the modulated sequence of user $p$ as ${\bf x}_p = [ x_{p}(1),...,x_{p}(n),...x_{p}(N)]$, where $x_{p}(n) = (-1)^{b_{p}(n)}$. In this paper, we concentrate on uplink transmissions over the quasi-static Rayleigh channel with BPSK modulation and assume that all users are perfectly aligned in time. The received signal can be represented as
\begin{equation} \label{equ:r_single}
 r(n) = \sum_{p=1}^{P}h_{p}x_{p}(n) + w(n),\quad n = 1, 2,..., N,
\end{equation}
where $h_{p}$ is the complex Gaussian channel gain of user $p$ and $w(n)$ is the additive Gaussian noise with variance $\sigma^2=\frac{N_0}{2}$ each dimension.  Both the amplitude of channel gain $|h_p|$ and the phase of the channel gain $\theta_p$ are useful in identifying the user.
\subsection{Multi-level Detection}
Let $P$ = 2 and $N$ = 1. Suppose that user $A$ and user $B$ share the same resource such as bandwidth, signature sequence and subcarrier allocation. 
The superimposed signal representing $b_A$ and $b_B$ transmitted from user $A$ and user $B$ respectively is denoted as $s_{AB}$, which should be one of the four levels given by $\{ S_{AB}(0), S_{AB}(1)$, $S_{AB}(2), S_{AB}(3)\}$. With given $h_A$ and $h_B$, the possible levels of $s_{AB}$ are shown in TABLE \ref{table:superimposed tx}. Note that a larger table can be established by a similar rule if more than two users share the same resource.
\begin{table}[H]
\caption{The superimposed signal and the corresponding data.}
	\begin{center}
		\begin{tabular}{ | >{\centering\arraybackslash}m{\columnwidth/12}  | >{\centering\arraybackslash}m{\columnwidth/10}  | >{\centering\arraybackslash}m{\columnwidth/10} | >{\centering\arraybackslash}m{\columnwidth/10} | >{\centering\arraybackslash}m{\columnwidth/10} | >{\centering\arraybackslash}m{\columnwidth/7} | }
		\hline
 		$i$ & $b_A$ & $b_B$ & $x_A$ & $x_B$ & $S_{AB}(i)$ \\ \hline
 		0 & 0 & 0 & 1 & 1 & $h_{A}+h_B$ \\ \hline 
 		1 & 0 & 1 & 1 & -1 & $ h_{A}-h_B$ \\ \hline 
 		2 & 1 & 0 & -1 & 1 & $-h_{A}+h_B$ \\ \hline 
 		3 & 1 & 1 & -1 & -1 & $-h_{A}-h_B$ \\ \hline
		\end{tabular}
		\label{table:superimposed tx}
	\end{center}
\end{table}

At the receiver, the received signal is $r$ = $r_{AB}$  = ${[h_{A}x_{A} + h_{B}x_{B}]} + {w}$.  The ${a \ posteriori}$ probability (APP) for $s_{AB}$ given the received signal $r_{AB}$ can be calculated by
\begin{align}  \label{equ:app}
	p_{i} &= \text{Pr}\left\{s_{AB} = S_{AB}(i)\middle|r_{AB}\right\} \notag \\
			  &= \text{Pr}\left\{r_{AB}\middle|s_{AB} = S_{AB}(i)\right\}\cdot\frac{\text{Pr}\left\{s_{AB} = S_{AB}(i)\right\}}{\text{Pr}\left\{r_{AB}\right\}} \notag 
\end{align}
where
\begin{align}
	\text{Pr}\left\{r_{AB}\middle|s_{AB} = S_{AB}(i)\right\} = \frac{1}{2\pi{\sigma}^{2}}\exp\left(-\frac{\lvert r_{AB}-S_{AB}(i)\rvert^{2}}{2{\sigma}^{2}}\right)
\end{align}
and $\text{Pr}\{s_{AB} = S_{AB}(i)\}$ is the $a \ priori$ probability for $s_{AB}$.
The APP values for ${b}_{A}$ and ${b}_{B}$ given the received signal $r_{AB}$ are respectively given by
\begin{equation}\left\{
	\begin{aligned}
	&\text{Pr}\left\{b_{A}=0 \middle| r_{AB}\right\}  
	=\ &\text{Pr}\left\{s_{AB} = S_{AB}(0)\middle|r_{AB}\right\} + \text{Pr}\left\{s_{AB} = S_{AB}(1)\middle|r_{AB}\right\}  
	=\ &p_{0} + p_{1} \\[1em]
	&\text{Pr}\left\{b_{A}=1\middle|r_{AB}\right\}  
	=\ &\text{Pr}\left\{s_{AB} = S_{AB}(2)\middle|r_{AB}\right\} + \text{Pr}\left\{s_{AB} = S_{AB}(3)\middle|r_{AB}\right\}  
	=\ &p_{2} + p_{3}
	\end{aligned}\right.
\end{equation}
and
\begin{equation}\left\{
	\begin{aligned}
	   &\text{Pr}\left\{b_{B}=0\middle|r_{AB}\right\}  
	=\ &\text{Pr}\left\{s_{AB} = S_{AB}(0)\middle|r_{AB}\right\} + \text{Pr}\left\{s_{AB} = S_{AB}(2)\middle|r_{AB}\right\}  
	=\ &p_{0} + p_{2} \\[1em]
	   &\text{Pr}\left\{b_{B}=1\middle|r_{AB}\right\} 
	=\ &\text{Pr}\left\{s_{AB} = S_{AB}(1)\middle|r_{AB}\right\} + \text{Pr}\left\{s_{AB} = S_{AB}(3)\middle|r_{AB}\right\}  
	=\ &p_{1} + p_{3}\ .
	\end{aligned}\right.
\end{equation}
The corresponding log likelihood ratio (LLR) values of bits $b_A$ and $b_B$ are
\begin{equation}  \label{equ:llr} 
	\begin{dcases}
e_{LLR}({b}_A) = \ln \frac {\text{Pr}\{b_{A}=0|r_{AB}\}} {\text{Pr}\{b_{A}=1|r_{AB}\}} = \ln\frac {p_{0}+p_{1}} {p_{2}+p_{3}}\\
e_{LLR}({b}_B) = \ln \frac {\text{Pr}\{b_{B}=0|r_{AB}\}} {\text{Pr}\{b_{B}=1|r_{AB}\}} = \ln\frac {p_{0}+p_{2}} {p_{1}+p_{3}}\ .
	\end{dcases}
\end{equation}
These LLR values can be used to estimate $b_A$ and $b_B$ by
\begin{align}
	\hat{b}_{p} = 
		\begin{dcases}
			0, &\text{if } e_{LLR}({b}_p)\geq 0\\
			1, &\text{otherwise}\\
		\end{dcases},\quad
		p=A\text{ or }B,
\end{align}	
or can be fed to a decoder if channel coding is considered. 

For $P$ = 3, the MLDT can be similarly executed, where the superimposed signal representing $b_A$, $b_B$ and $b_C$ transmitted from users $A$, $B$ and $C$ respectively should be one of the eight levels given by $\{S_{ABC}(0), S_{ABC}(1)$, $S_{ABC}(2), S_{ABC}(3), S_{ABC}(4), S_{ABC}(5)$, $S_{ABC}(6), S_{ABC}(7)\}$, where $S_{ABC}(i) = (-1)^{b_{A}}h_{A} + (-1)^{b_{B}}h_{B} + (-1)^{b_{C}}h_{C}$ and $i$ = $4b_A$ + $2b_B$ + $b_C$.
\subsection{BER Analysis}
By assuming equally likely $a \ priori$ probability for each $s_{AB}$ (or $s_{ABC}$), the bounds of BER for MLDT receiver with $P$ = 2 (or 3) respectively over independent and identically distributed (i.i.d.) Rayleigh fading channels can be derived as follows.  Let
\begin{equation}
p_{R}(r) = \frac{2r}{\mathbb{E}[R^2]}\exp\left(-\frac{r^2}{\mathbb{E}\left[R^2\right]}\right)
\end{equation}
be the probability density function of a Rayleigh distributed random variable $R$.\\
Let ${\displaystyle Q(x)=\frac{1}{\sqrt{2\pi}}\int_x^\infty\exp\left(-\frac{u^2}{2}\right)\mathrm{d}u}$ and  let
$\overline{\gamma}$ = $\frac{1}{N_0}\mathbb{E}\left[\abs{h_A}^2\right]$ = $\frac{1}{N_0}\mathbb{E}\left[\abs{h_B}^2\right]$ = $\frac{1}{N_0}\mathbb{E}\left[\abs{h_C}^2\right]$, where $|h_A|$, $|h_B|$ and $|h_C|$ are i.i.d. Rayleigh random variables.
\subsubsection {$P = 1$}
It is well known that the average BER for the BPSK modulation over the Rayleigh fading channel \cite{Goldsmith05} is
\begin{equation}\label{equ:uncoded_lowerbound}
	\bar{P_e} = \int\limits_0^{\infty} Q\left(\frac{2r}{\sqrt{2N_0}}\right) p_{R}(r)\mathrm{d}r=\frac{1}{2}\left(1-\sqrt{\frac{\overline{\gamma}}{1+\overline{\gamma}}}\right).
\end{equation}

\subsubsection {$P = 2$}
The BER of user $A$ is
\begin{align} \label{equ:pe2}
	 P_e 
	 = \ &\frac{1}{2}\text{Pr}\left\{\hat{b}_{A}=1\middle|b_{A}=0\right\}+\frac{1}{2}\text{Pr}\left\{\hat{b}_{A}=0\middle|b_{A}=1\right\} \notag\\
	=\ &\text{Pr}\left\{\hat{b}_{A}=1\middle|b_{A}=0\right\} \ (\mbox{from symmetry})\notag\\
	=\ &\frac{1}{2}\text{Pr}\left\{\hat{b}_{A}=1\middle|s_{AB} = S_{AB}(0)\right\} 
	 +\frac{1}{2}\text{Pr}\left\{\hat{b}_{A}=1\middle|s_{AB} = S_{AB}(1)\right\}.
\end{align}

The Euclidean distance between $S_{AB}(0)$ and $S_{AB}(i)$ is $2|h_A|$ for $i$ = 2,  and is $2|h_A+h_B|$ for $i$ = 3.   Moreover, the Euclidean distance between $S_{AB}(1)$ and $S_{AB}(i)$ is $2|h_A-h_B|$ for $i$ = 2,  and is $2|h_A|$ for $i$ = 3.
Hence, an upper bound of $P_e$ can be derived as
\begin{align} \label{equ:up_last2a}
	 P_e 
	&\leq\ \frac{1}{2}\Bigg\{\sum_{i\in\{2,3\}}\text{Pr}\left\{\abs{r_{AB}-S_{AB}(i)}<\abs{r_{AB}-S_{AB}(0)}\middle|S_{AB}(0)\right\} \Bigg\} \notag\\	  
	 &+ \frac{1}{2}\Bigg\{\sum_{i\in\{2,3\}}\text{Pr}\left\{\abs{r_{AB}-S_{AB}(i)}<\abs{r_{AB}-S_{AB}(1)}\middle|S_{AB}(1)\right\} \Bigg\}
\end{align}
Assume that $|h_A|$ and $|h_B|$ are independent and identically Rayleigh distributed.  Also assume that $\theta_A$ and $\theta_B$ are independent and identically uniformly distributed.  The average of the righthand side of (\ref{equ:up_last2a}) is
\begin{align} \label{equ:up_last2b}
	 \ &\frac{1}{2}\int\limits_{0}^{\infty}\int\limits_{0}^{\infty}\int\limits_0^{\frac{\pi}{2}}\int\limits_0^{\frac{\pi}{2}}\left\{Q\left(\frac{2\abs{h_A}}{\sqrt{2N_0}}\right)+Q\left(\frac{2\abs{h_A+h_B}}{\sqrt{2N_0}}\right)\right\} \cdot p_{R}(|h_{B}|)p_{R}(|h_{A}|)\left(\frac{2}{\pi}\right)^2\mathrm{d}|h_{B}|\mathrm{d}|h_{A}|\mathrm{d}\theta_{B}\mathrm{d}\theta_{A} \notag\\ 
	+ & \frac{1}{2}\int\limits_{0}^{\infty}\int\limits_{0}^{\infty}\int\limits_0^{\frac{\pi}{2}}\int\limits_0^{\frac{\pi}{2}}\left\{Q\left(\frac{2\abs{h_A}}{\sqrt{2N_0}}\right)+Q\left(\frac{2\abs{h_A-h_B}}{\sqrt{2N_0}}\right)\right\} \cdot
p_{R}(|h_{B}|)p_{R}(|h_{A}|)\left(\frac{2}{\pi}\right)^2\mathrm{d}|h_{B}|\mathrm{d}|h_{A}|\mathrm{d}\theta_{B}\mathrm{d}\theta_{A}.
\end{align}
\newpage
Since $|h_A+h_B|$ and $|h_A-h_B|$ are both Rayleigh distributed with variance twice of that for $|h_A|$, the average upper bound of BER provided in (\ref{equ:up_last2b}) can be simplified as
\begin{equation} \label{equ:uncoded_upperbound2}
\bar{P}e \le \frac{1}{2}\left(1-\sqrt{\frac{\overline{\gamma}}{1+\overline{\gamma}}}\right) + \frac{1}{2}\left(1-\sqrt{\frac{2\overline{\gamma}}{1+2\overline{\gamma}}}\right)
	\end{equation}
Since the BER of the collided users cannot be lower than that of the non-collided users, the average BER for the BPSK modulation over Rayleigh fading channels can be regarded as the lower bound of the MLDT receiver.  We have
\begin{equation}\label{equ:uncoded_lowerbound}
	\bar{P}_e \geq \frac{1}{2}\left(1-\sqrt{\frac{\overline{\gamma}}{1+\overline{\gamma}}}\right).
\end{equation}
\begin{figure}[H]
		\centering
		\includegraphics[width=16cm, height=10cm]{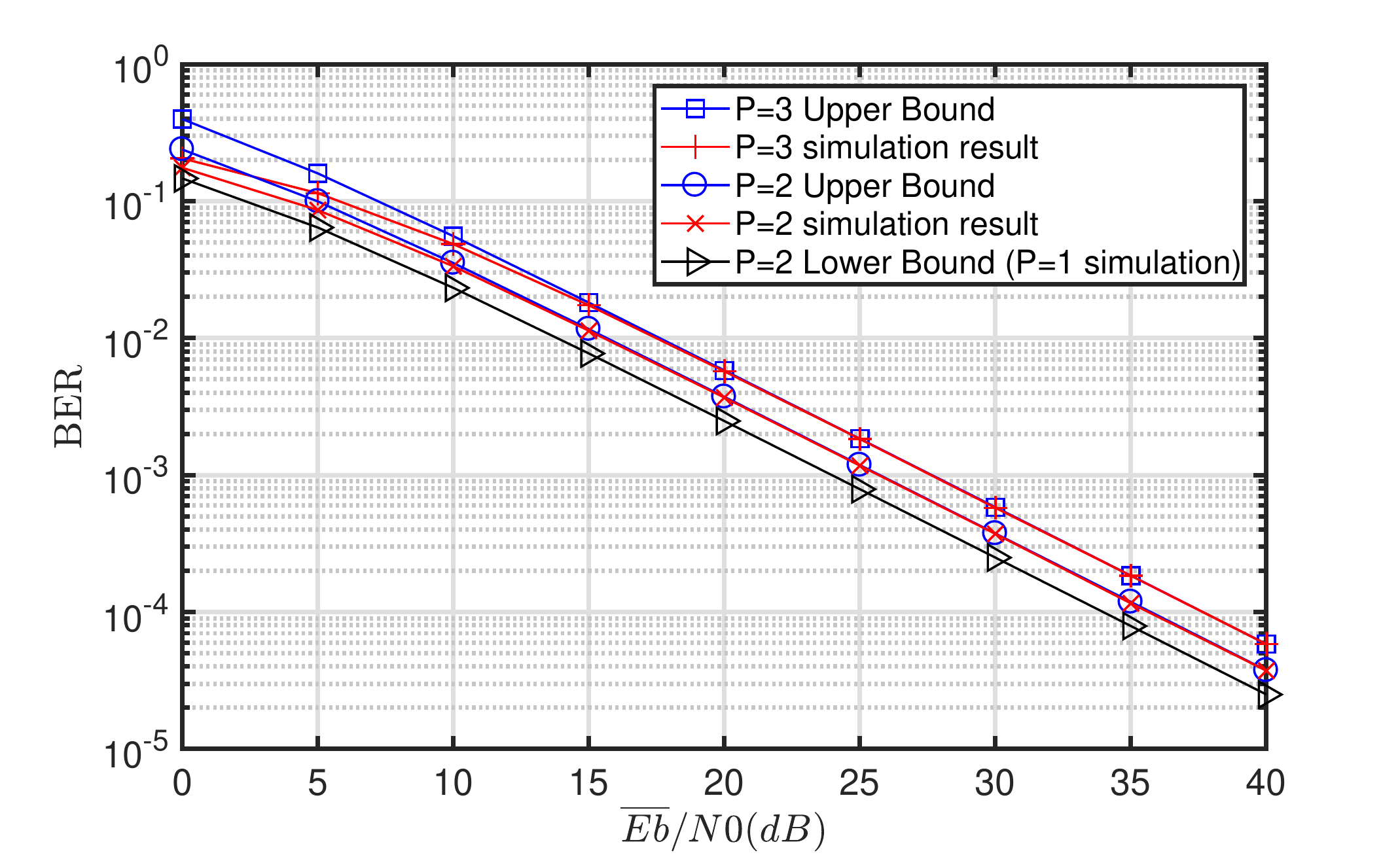}
		\caption{Bounds of BER and simulated results for the uncoded MLDT receiver over Rayleigh fading channels.}
		\label{fig:analysis_uncoded}
\end{figure}
\subsubsection {$P = 3$}
Similar to (\ref{equ:pe2}), (\ref{equ:up_last2a}),  (\ref{equ:up_last2b}) and (\ref{equ:uncoded_upperbound2}) for $P$ = 2, upper bounds of $P_e$ and $\bar{P}_e$ for $P$ = 3 can be derived as,
\begin{align} \label{equ:up_last3a}
	 P_e 
	&\leq\ \frac{1}{4}\sum_{i \in \{4,5,6,7\}}\text{Pr}\left\{\abs{r_{AB}-S_{AB}(i)}<\abs{r_{AB}-S_{AB}(0)}\middle|S_{AB}(0)\right\} \notag \\	  
	 &+ \frac{1}{4}\sum_{i \in \{4,5,6,7\}} \text{Pr}\left\{\abs{r_{AB}-S_{AB}(i)}<\abs{r_{AB}-S_{AB}(1)}\middle|S_{AB}(1)\right\} \notag \\
	 &+ \frac{1}{4}\sum_{i \in \{4,5,6,7\}} \text{Pr}\left\{\abs{r_{AB}-S_{AB}(i)}<\abs{r_{AB}-S_{AB}(2)}\middle|S_{AB}(2)\right\} \notag \\
	 &+ \frac{1}{4}\sum_{i \in \{4,5,6,7\}} \text{Pr}\left\{\abs{r_{AB}-S_{AB}(i)}<\abs{r_{AB}-S_{AB}(3)}\middle|S_{AB}(3)\right\} 
\end{align}
and
\begin{align} \label{equ:up_last3b} 
	\bar{P}_e &\le \frac{1}{2}\left(1-\sqrt{\frac{\overline{\gamma}}{1+\overline{\gamma}}}\right) + \left(1-\sqrt{\frac{2\overline{\gamma}}{1+2\overline{\gamma}}}\right)
	+ \frac{1}{2}\left(1-\sqrt{\frac{3\overline{\gamma}}{1+3\overline{\gamma}}}\right)
\end{align}


\subsubsection {Numerical Results}
For a large $x$, the term $1-\sqrt{x/(1+x)}$ can be approximated by $1/(2x)$.  Hence, for $P=1$, we have $\bar{P}_e \approx 1/(4\overline{\gamma})$; for $P=2$, we have $\bar{P}_e \approx 3/(8\overline{\gamma})$; for $P=3$, we have $\bar{P}_e \approx 7/(12\overline{\gamma})$. Hence, we expect that for high SNR, the penalty for doubling the user number is the increase of BER by about 50\% and for tripling the user number is the increase of BER by about 133\%.   Fig. \ref{fig:analysis_uncoded} shows the analytical results obtained in (\ref{equ:uncoded_upperbound2}) (\ref{equ:uncoded_lowerbound}) (\ref{equ:up_last3b}) and the simulation results, where $\overline{E}_b/N_0$ denote the average of $E_b/N_0$ over the Rayleigh fading channel. We see that simulation results are very close to the analytic prediction.

\section{Coded MLDT} \label{sec:coded MLDT}  
\begin{figure}[H]
	\centering
	\includegraphics[width=16cm, height=3cm]{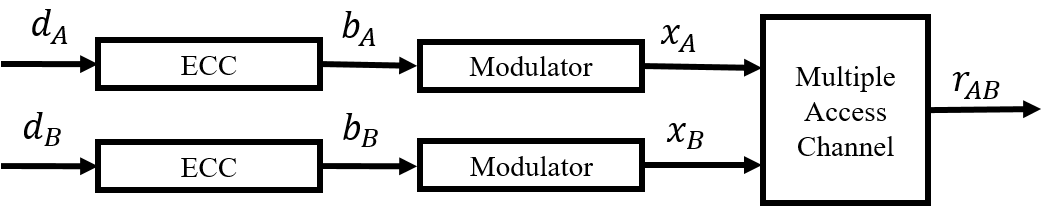}
	\caption{The transmitters of a coded system with $P$ = 2.}
	\label{fig:ldpc_cdma_tx}
\end{figure}

Consider a symbol-synchronous coded system in which $P$ users transmit messages over independent quasi-static Rayleigh channels. For user $p$, a message block ${\bf d}_{p}$ is encoded by the encoder of an error-correcting code $V$ into a codeword (or code sequence) ${\bf b}_{p}$.  We will first consider the design for which the MLDT receiver is followed by multiple SPA (sum product algorithm) decoders and then consider the design for which the MLDT receiver is followed by a single GSPA (generalized sum product algorithm)  \cite{PLNC} decoder.  The transmitters with $P$ = 2 is illustrated in Fig. \ref{fig:ldpc_cdma_tx}.

\subsection{MLDT Receiver using SPA}
\begin{figure}[H]
	\centering
	\includegraphics[width=16cm, height=4.5cm]{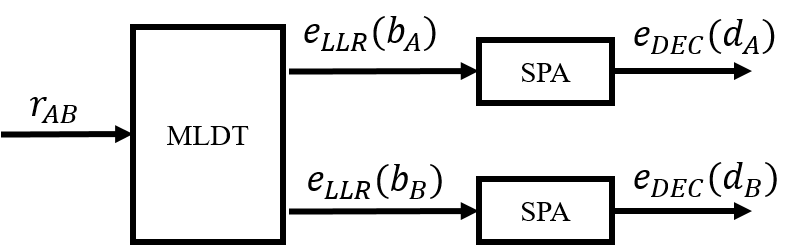}
	\caption{The MLDT receiver followed by two SPA decoders for a coded system with $P$ = 2.}
	\label{fig:ldpc_mldt_rx}
\end{figure}

For the coded system with $P$ = 2, the MLDT receiver followed by two conventional SPA decoders is illustrated in Fig. \ref{fig:ldpc_mldt_rx}.
The LLR values $e_{LLR}(b_{A})$ and $e_{LLR}(b_{B})$ obtained from the MLDT receiver are respectively fed to two binary decoders which use the SPA to obtain the decoded LLR values $e_{DEC}(d_{A})$ and $e_{DEC}(d_{B})$ for message bits $d_A$ and $d_B$ respectively.

\subsection{MLDT Receiver using GSPA}
\begin{figure}[H]
	\centering
	\includegraphics[width=16cm, height=5.5cm]{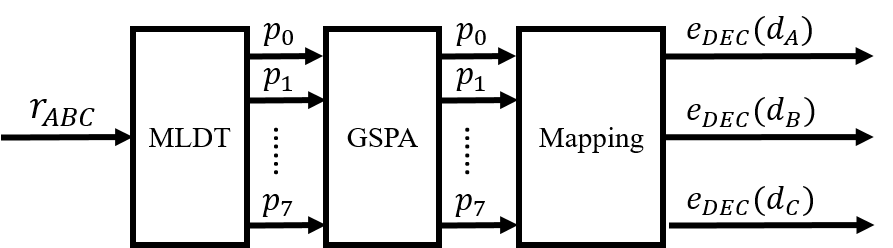}
	\caption{The MLDT receiver followed by a GSPA decoder for a coded system with $P$ = 3.}
	\label{fig:ldpc_plnc_rx}
\end{figure}

For the coded CDMA system, the MLDT receiver followed by $P$ SPA decoders can be replaced by an MLDT receiver followed by a single generalized sum product algorithm (GSPA) decoder to obtain the decoded LLR values $e_{DEC}(d_{1})$, $e_{DEC}(d_{2}),...$ and $e_{DEC}(d_{P})$ for message bits $d_1$, $d_2,...$ and $d_P$ respectively.  Such a receiver with $P$ = 3 is depicted in Fig. \ref{fig:ldpc_plnc_rx}.
 Here, the ECC code $V$ is a binary one.  However, both the codewords represented by code bits $\{b_{A}\}$, $\{b_{B}\}$ and $\{b_{C}\}$ for users $A$ , $B$ and $C$  respectively must satisfy the same factor graph of code $V$.  A variable node in the conventional SPA decoder of a binary code is represented by the bit $b_k$. Such a variable node in the GSPA decoder is now represented by a three-tuple $(b_{A}, b_{B}, b_{C})$ for which its likelihood vector is ${\bf p}$ = $\{p_0, p_1, p_2, p_3, p_4, p_5, p_6, p_7\}$, where $p_i$ at the input of the GSPA decoder is defined in a way similar to (\ref{equ:app}) through replacing $r_{AB}$, $s_{AB}$ and $S_{AB}$ by  $r_{ABC}$, $s_{ABC}$ and $S_{ABC}$ respectively.
 
The updated likelihood vector ${\bf p}$ of the GSPA in the iterative operation can be obtained as follows. 
Suppose that there is a degree-3 variable node which takes two 
likelihood vectors 
${\bf p}$ = $\left[p_0\ p_1\ p_2\ p_3\ p_4\ p_5\ p_6\ p_7\right]$ and ${\bf q}$ = $\left[q_0\ q_1\ q_2\ q_3\ q_4\ q_5\ q_6\ q_7\right]$ as input.  The updated likelihood vector will be 
\begin{equation}
	\text{VAR}({\bf p},{\bf q})=\beta\left[p_0q_0\ p_1q_1\ p_2q_2\ p_3q_3\ p_4q_4\ p_5q_5\ p_6q_6\ p_7q_7\right],
\end{equation}
where $\beta$ is a normalized factor. Suppose that there is a degree-3 check node which takes two 
likelihood vectors ${\bf p}$ and ${\bf q}$ as input.  The updated likelihood vector will be
\begin{equation}
	\text{CHK}({\bf p},{\bf q})= \left[
	\begin{split}
	&p_0q_0+p_1q_1+p_2q_2+p_3q_3+p_4q_4+p_5q_5+p_6q_6+p_7q_7\\
	&p_0q_1+p_1q_0+p_2q_3+p_3q_2+p_4q_5+p_5q_4+p_6q_7+p_7q_6\\
	&p_0q_2+p_1q_3+p_2q_0+p_3q_1+p_4q_6+p_5q_7+p_6q_4+p_7q_5\\
	&p_0q_3+p_1q_2+p_2q_1+p_3q_0+p_4q_7+p_5q_6+p_6q_5+p_7q_4\\
	&p_0q_4+p_1q_5+p_2q_6+p_3q_7+p_4q_0+p_5q_1+p_6q_2+p_7q_3\\
	&p_0q_5+p_1q_4+p_2q_7+p_3q_6+p_4q_1+p_5q_0+p_6q_3+p_7q_2\\
	&p_0q_6+p_1q_7+p_2q_4+p_3q_5+p_4q_2+p_5q_3+p_6q_0+p_7q_1\\
	&p_0q_7+p_1q_6+p_2q_5+p_3q_4+p_4q_3+p_5q_2+p_6q_1+p_7q_0
	\end{split}
	\right]^T.
\end{equation}
For a node with degree more than 3, the update likelihood vector can be extended by 
\begin{equation}
	\text{VAR}({\bf p},{\bf q},\cdots)=\text{VAR}({\bf p},\text{VAR}({\bf q},\text{VAR}(\cdot,\cdot))),
\end{equation}
	and
\begin{equation}
	\text{CHK}({\bf p},{\bf q},\cdots)=\text{CHK}({\bf p},\text{CHK}({\bf q},\text{CHK}(\cdot,\cdot))).
\end{equation}

After the iterations, we can determine the LLR values of the collided users by

\begin{equation}
\left\{
\begin{aligned}
	&e_{DEC}({d}_{A}) =
\ln\frac {p_{0}+p_{1}+p_{2}+p_{3}} {p_{4}+p_{5}+p_{6}+p_{7}} \\
&e_{DEC}({d}_{B}) =
 \ln\frac {p_{0}+p_{1}+p_{4}+p_{5}} {p_{2}+p_{3}+p_{6}+p_{7}}\ . \\
&e_{DEC}({d}_{C}) =
 \ln\frac {p_{0}+p_{2}+p_{4}+p_{6}} {p_{1}+p_{3}+p_{5}+p_{7}}
\end{aligned}
\right.
\end{equation}
The $P$ = 1 and $P = 2$ scenarios are simply the degenerate cases of $P$ = 3 case. 

\subsection{LDPC-Coded System}
\begin{figure}[H]
	\centering
	\includegraphics[width=16cm, height=9cm]{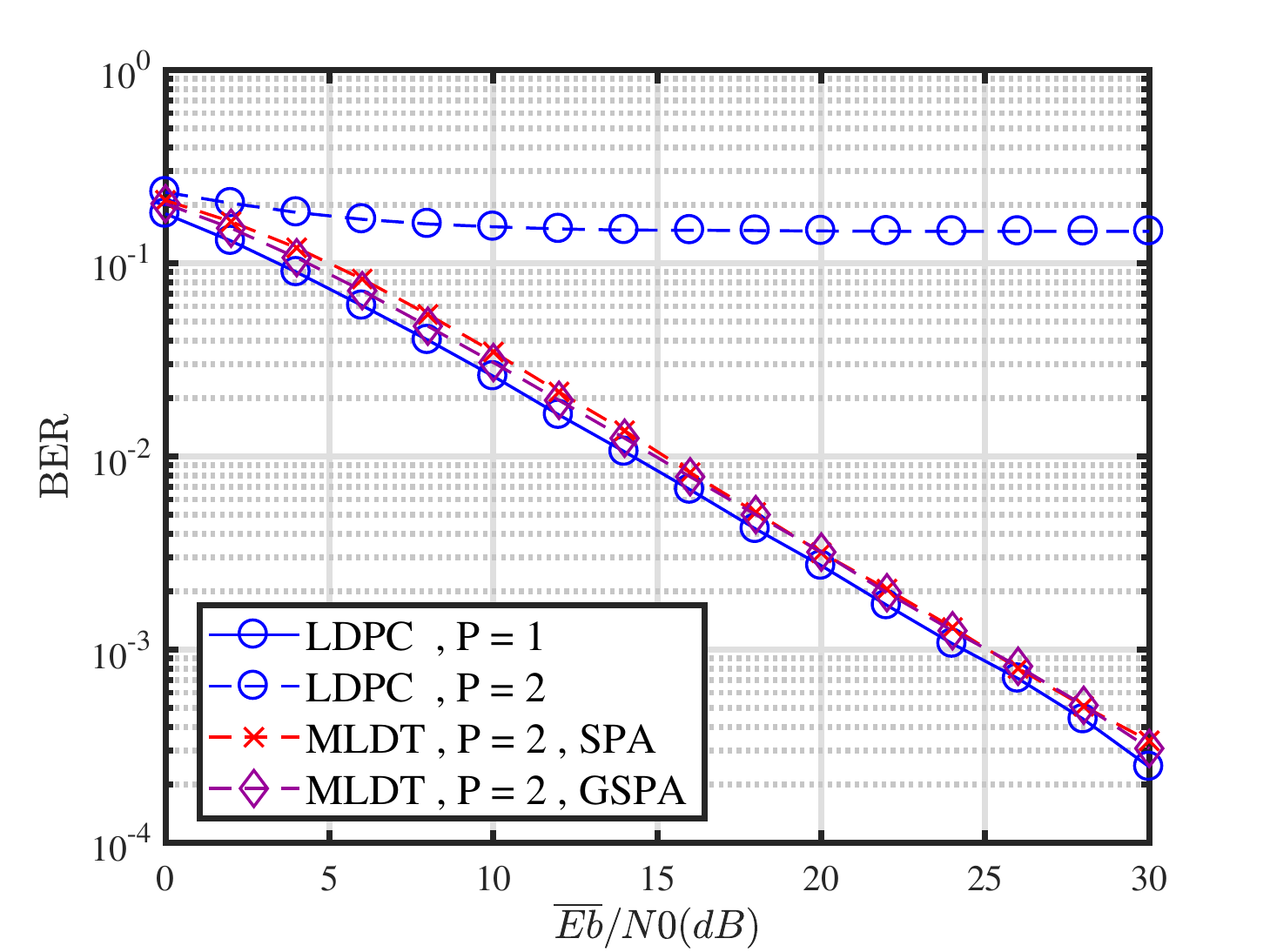}%
	\caption{LDPC-coded system over single-path Rayleigh fading channels.}
	\label{fig:sim_LDPC_CDMA}
\end{figure}

Fig. \ref{fig:sim_LDPC_CDMA} shows the BER performances of an LDPC-coded system with $P$ = 2, where the (1008, 504) $(3,6)$-regular LDPC code \cite{Mackay} is used.  The single-path Rayleigh fading channel is considered. The number of iterations for both SPA and GSPA is set to 10. The ``LDPC, P = 2" curve shows the BER performances of the conventional receiver for two collided users without MLDT.  Clearly, such a arrangement has extremely poor performances.  With the MLDT structure for $P$ = 2, using either the two SPA decoders or the single GSPA decoder can achieve similar performances which are only slightly inferior to the $P$ = 1 case. 

Compared to the uncoded MLDT system, the LDPC coded MLDT system does not provide noticeable coding gain.  Hence, in the following section, we will conduct the capacity analysis to see whether we can achieve benefit by applying ECC to the MLDT system.

\subsection{Capacity Analysis}
It would be interesting to compare capacity $C_{AB,BPSK,R}$ of two users employing BPSK averaged over independent quasi-static Rayleigh fading channels to the capacity $C_{QPSK,R}$ of a single user employing QPSK averaged over the Rayleigh fading channel.   
Let $x_A, x_B \in \{+1,-1\}$, with fixed channel gain of $h_A$ and $h_B$, the superimposed signal without AWGN will be $s_{AB}$ = $x_Ah_A+x_Bh_B$ = $(x_A|h_A|+x_B|h_B|\cos\theta)+j(x_B|h_B|\sin\theta)$.

Assume equally likely $a \ priori$ probability for $s_{AB}$. Then, we have 
\begin{align} \label{equ:cab}
C_{AB,BPSK,R} = &\sum_{x_A,x_B} \frac{1}{4}\int\limits_{-\infty}^{\infty}\int\limits_{-\infty}^{\infty}\int\limits_0^{2\pi}\int\limits_{0}^{\infty}\int\limits_{0}^{\infty}f(x_A,x_B,u,v,\theta,|h_A|,|h_B|) \notag \\
&\log \left\{f(x_A,x_B,u,v,\theta,|h_A|,|h_B|)  \big[\frac{1}{4}{\sum_{x_A,x_B}f(x_A,x_B,u,v,\theta,|h_A|,|h_B|) }\big]^{-1}\right\} \notag \\  
&\cdot p_{R}(|h_{A}|)p_{R}(|h_{B}|)\frac{1}{2\pi}\mathrm{d}u\mathrm{d}v\mathrm{d}\theta\mathrm{d}|h_{A}|\mathrm{d}|h_{B}| 
\end{align}
where 
\begin{align} \label{equ:cabf}
&f(x_A,x_B,u,v,\theta,|h_A|,|h_B|) = \frac{1}{2\pi\sigma^2}\exp\left\{\frac{-|r_{AB}-s_{AB}|^2}{2\sigma^2}\right\}\notag\\ 
&=\frac{1}{2\pi\sigma^2}\exp\left\{\frac{-((u-(x_A|h_A|+x_B|h_B|\cos\theta))^2 + (v-x_B|h_B|\sin\theta)^2)}{2\sigma^2}\right\},
\end{align}
 and $r_{AB}=u+jv$.
The capacity $C_{QPSK,R}$ can be obtained by setting $|h_A|$ = $|h_B|$ and $\theta = \pi/2$ in (\ref{equ:cab})  and (\ref{equ:cabf}).

Let $\overline{E}_s/N_0$ denote the average of $E_s/N_0$ over the Rayleigh fading channel.
From Fig. \ref{fig:Capacity Compare},
we see that under the quasi-static Rayleigh fading scenarios, $C_{AB,BPSK,R}$ is close to $C_{QPSK,R}$, especially in the high SNR regime.  The capacity of $C_{QPSK,A}$ over the AWGN channel is also provided in Fig. \ref{fig:Capacity Compare} as a reference for comparison.  This result implies that in the coded systems over the Rayleigh fading channels, MLDT for two-user multiple access employing BPSK suffers only a slight loss of average capacity as compared to a single user employing QPSK transmission.

\subsection{Raptor-Coded System}
In Fig. \ref{fig:sim_LDPC_CDMA}, we note that over the single-path Rayleigh fading channel, using LDPC virtually obtains no coding gain as compared to the uncoded system.  This is probably due to the fact that a fixed-rate ECC cannot cope with the varying SNR in the deep fading condition. In case that feedback channel is available, Raptor coding \cite{Shokrollahi}  can indefinitely increase its redundancy until the decoding is successful.  A Raptor code $V$ can be constructed as the concatenation of a high-rate ECC $V'$ followed by a rateless Luby Transform (LT) code $C$ which can generate  limitless output stream ${\bf b}_{p}$ until the transmitter receives an "ACK" signal sent by the receiver through the feedback channel.  Hence, the length $N$ of ${\bf b}_{p}$ is a random variable.  The system throughput will be 
\begin{equation}
R_{V} = \frac{k}{\mathbb{E}\left[N\right]},
\end{equation}
where $k$ is the length of each message block  ${\bf d}_{p}$.

The transmitters of a Raptor coded system with $P$ = 2 can also be illustrated in Fig. \ref{fig:ldpc_cdma_tx}, where the the ECC encoder $V$ is now a Raptor encoder.
For Raptor coded MLDT system with $P$ = 2,  both user $A$ and user $B$ will increase the output stream length $N$  until both ${\bf b}_{A}$ and ${\bf b}_{B}$ are successfully recovered. 

\begin{figure}[H]
	\centering
	\includegraphics[width=16cm, height=3cm]{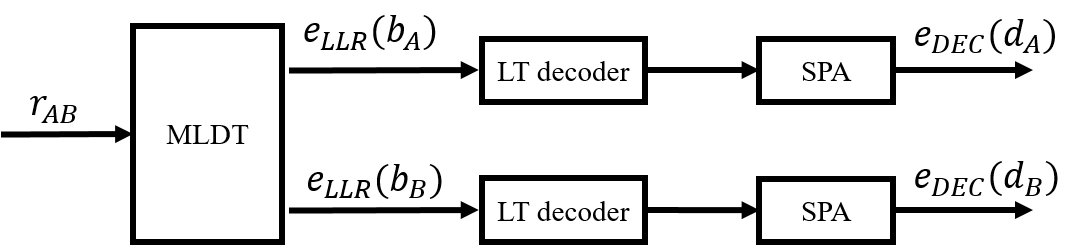}%
	\caption{The MLDT receiver followed by two LT decoders and two SPA decoders for an Raptor-coded system. $P$ = 2.}
	\label{fig:raptor_MLDT_rx_SPA}
\end{figure}

The MLDT receiver with $P$ = 2 using multiple SPA decoders  illustrated in Fig. \ref{fig:ldpc_mldt_rx} must be modified as shown in Fig.~\ref{fig:raptor_MLDT_rx_SPA}, where each SPA decoder is used as the decoder for the ECC $V'$ which is usually a binary LDPC code.

\begin{figure}[H]
	\centering
	\includegraphics[width=17cm, height=4cm]{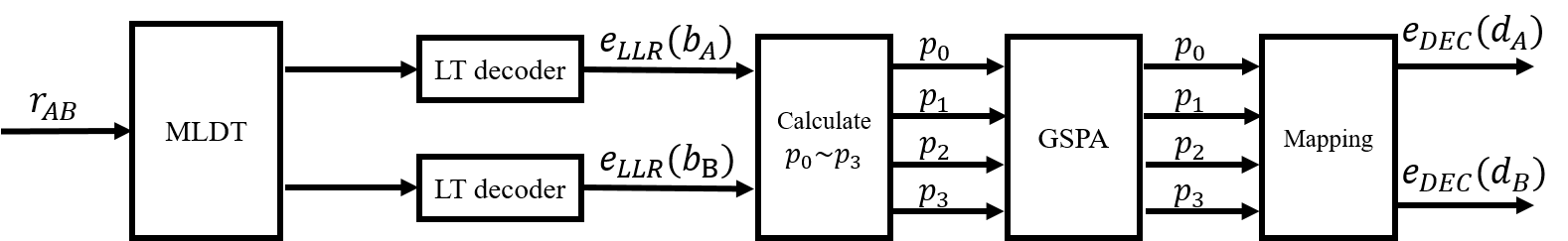}%
	\caption{The MLDT receiver followed by two LT decoders and a single GSPA decoder for an Raptor-coded system. $P$ = 2.}
	\label{fig:raptor_MLDT_rx_GSPA}
\end{figure}

Likewise, the MLDT receiver with $P$ = 2 using a single GSPA decoder must be modified as shown in Fig.~\ref{fig:raptor_MLDT_rx_GSPA}, where the single GSPA decoder is used for the decoding of the two binary ECC $V'$.
LLR values $e_{LLR}(b_{A})$ and $e_{LLR}(b_{B})$ obtained from the MLDT receiver and LT decoders will be used to calculate likelihood vector $(p_0, p_1, p_2, p_3)$ according to (\ref{equ:raptorcalculate_p}).   This likelihood vector $(p_0, p_1, p_2, p_3)$ is then  fed to a GSPA decoder.

\begin{equation}\label{equ:raptorcalculate_p}
\begin{dcases}
&p_0 = \beta(1-\frac{1}{1+\exp[e_{LLR}(b_{A})]})(1-\frac{1}{1+\exp[e_{LLR}(b_{B})]})\\
&p_1 = \beta(1-\frac{1}{1+\exp[e_{LLR}(b_{A})]})(\frac{1}{1+\exp[e_{LLR}(b_{B})]})\\
&p_2 = \beta(\frac{1}{1+\exp[e_{LLR}(b_{A})]})(1-\frac{1}{1+\exp[e_{LLR}(b_{B})]})\\
&p_3 = \beta(\frac{1}{1+\exp[e_{LLR}(b_{A})]})(\frac{1}{1+\exp[e_{LLR}(b_{B})]}),
\end{dcases}
\end{equation}
where $\beta$ is the normalization factor.

\begin{figure}[H]
		\centering
		\includegraphics[width=16cm, height=10cm]{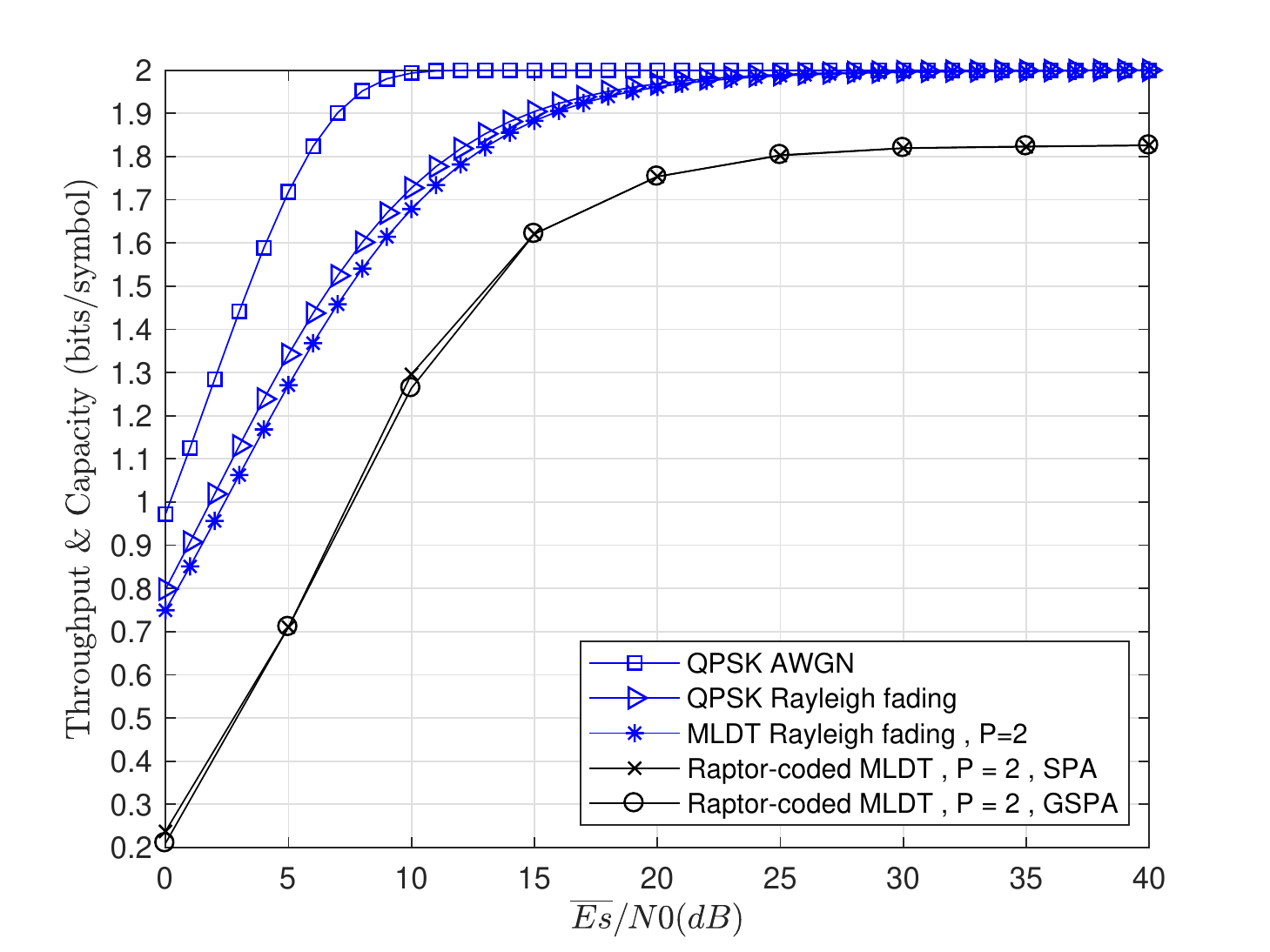}
		\caption{Throughput performances of Raptor-coded system and the derived capacities of QPSK and MLDT}
		\label{fig:Capacity Compare}
\end{figure}

The throughput performances of a Raptor coded MLDT system are shown in 
Fig. \ref{fig:Capacity Compare}, where the Raptor code is designed in   
\cite{Kuo}, of which
the ECC $V'$ is a (10000,9500) binary LDPC code and the LT code has an adaptive degree distribution.   In the simulation, each incremental redundancy (IR) contains 400 symbols, the number of iterations for LT decoder is set to 200  and the number of iterations for both SPA and GSPA is set to 100.   For each block if Raptor-code rate less than $\frac{1}{4}$ and a certain user is not able to obtain successful decoding, we stop decoding and the associated $N$ output bits will contribute to zero message bit in calculating the throughput.  Although Raptor-coded systems performance is about 10\% to 66\% below the associated capacity, the Raptor coded MLDT system does achieve appreciable coding gains in the quasi-static environment. Raptor code designed for low SNR \cite{Jayasooriya17} over the AWGN channel may also help to increase the throughput of MLDT system for low SNR over the independent quasi-static Rayleigh fading channels.

\section{MLDT for CDMA Systems} \label{sec:MLDT_CDMA}
Consider a symbol-synchronous uncoded CDMA system with $KP$ users. For user $(k,p)$, $1 \le k \le K$, $1 \le p \le P$, let $b_{k,p}$ be the transmitted message bit and ${\bf s}_k=[s_{k}(1),...,s_{k}(j),...,s_{k}(J)]$ be the spreading signature with spreading length $J$. Denote the modulated sequence of user $(k,p)$ as ${\bf x}_{k,p}$ = $[x_{k,p}(1),...,x_{k,p}(j),...,x_{k,p}(J)]$, where $x_{k,p}(j)=b_{k,p}s_{k}(j)$.

\begin{figure}[H]
	\centering
	\includegraphics[width=16cm, height=8cm]{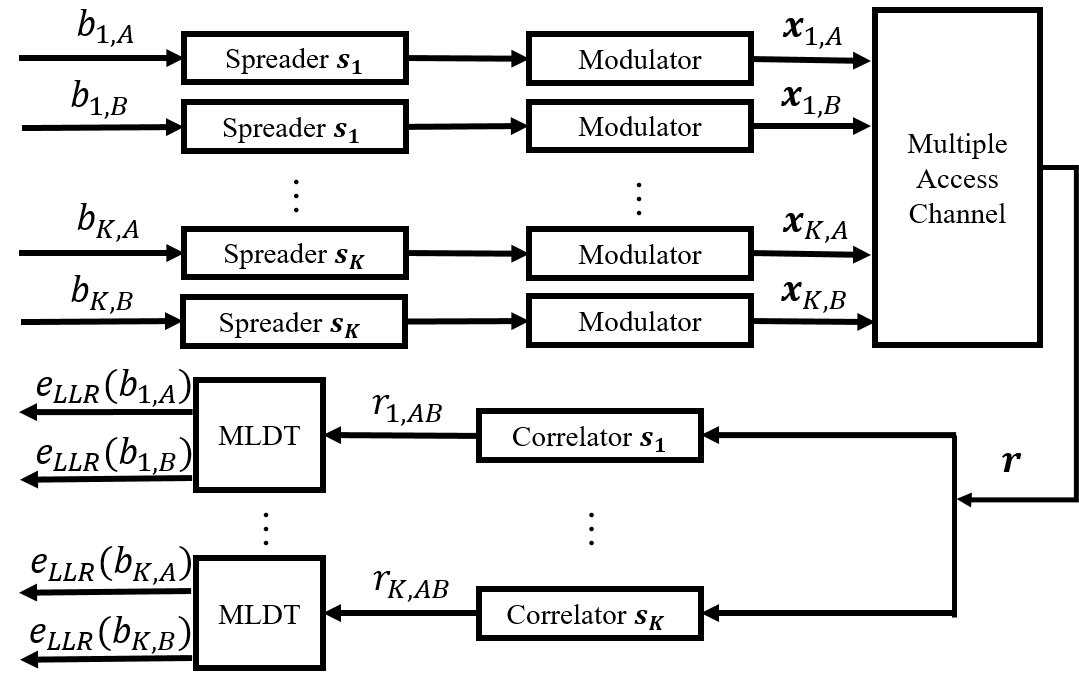}
	\caption{Illustration of an uncoded CDMA system with MLDT receivers. $P$ = 2.}
	\label{fig:uncoded MLDT}
\end{figure}

We consider the multipath channels which have $L$ paths, where the delay between adjacent paths is the period of a chip and all the paths have equal average powers. The received signal is
\begin{equation} \label{equ:r_single}
 r(j) = \sum_{\ell=0}^{L-1}\sum_{p=1}^{P}\sum_{k=1}^{K}h_{k,p,\ell}x_{k,p}(j-\ell) + w(j),\quad j = 1,...,J+L-1,
\end{equation}
where $h_{k,p,\ell}$ is the channel coefficient of $\ell$-th path for user $(k,p)$ and $x_{k,p}(j)$ with $j \notin \{1, 2, \cdots, J\}$ is modulated from a message bit $b_{k,p}$.  We assume that all the $h_{k,p,\ell}$ are independently identical and Rayleigh distributed.
The uncoded CDMA system with user collisions and MLDT receiver for $P$ = 2 is depicted in Fig. \ref{fig:uncoded MLDT}. 

For $P=2$, with correlator for ${\bf s}_k$, we have
\begin{align}  \label{equ:bab_single}
	r^{l}_{k,AB}
	&= \frac 1 J\sum_{j=1}^{J}r(j+l)s_{k}(j) \notag \\
	&= \frac 1 J\sum_{j=1}^{J}{[h_{k,A,l}x_{k,A}(j) + h_{k,B,l}x_{k,B}(j)]}s_{k}(j) + w_{ms} + w_{mp} + {w}_k,
\end{align}
where $l=0,1,...L-1, w_{ms} = \frac{1}{J}\sum_{j=1}^{J}\sum_{\ell=0}^{L-1}\sum_{k'\neq k} [h_{k',A,\ell}x_{k',A}(j+l-\ell)+h_{k',B,\ell}x_{k',B}(j+l-\ell)]s_{k}(j)$ is the interference from users employing sequences ${\bf s}_{k'} \neq {\bf s}_k$, ${w}_{mp} = \frac{1}{J}\sum_{j=1}^{J}\sum_{\ell\neq l}[h_{k,A,\ell}x_{k,A}(j+l-\ell)+h_{k,B,\ell}x_{k,B}(j+l-\ell)]s_{k}(j)$ is the interference from other paths, and ${w}_k$ = $\frac 1 J\sum_{j=1}^{J}w(j)s_{k}(j)$ is the Gaussian distributed noise with zero mean and variance ${\sigma}^2/{J}$. 
The parameters $p_i$, $e_{LLR}({b}_{k,A})$, $e_{LLR}({b}_{k,B})$ should be modified to $p^{l}_i$, $e^l_{LLR}({b}_{k,A})$, $e^l_{LLR}({b}_{k,B})$ respectively.  Then, we have 
\begin{equation}  \label{equ:llr2} 
	\begin{dcases}
e^l_{LLR}({b}_{k,A}) = \ln \frac {\text{Pr}\{b_{k,A}=0|r^l_{k,AB}\}} {\text{Pr}\{b_{k,A}=1|r^l_{k,AB}\}} = \ln\frac {p^l_{k,0}+p^l_{k,1}} {p^l_{k,2}+p^l_{k,3}}\\
e^l_{LLR}({b}_{k,B}) = \ln \frac {\text{Pr}\{b_{k,B}=0|r^l_{k,AB}\}} {\text{Pr}\{b_{k,B}=1|r^l_{k,AB}\}} = \ln\frac {p^l_{k,0}+p^l_{k,2}} {p^l_{k,1}+p^l_{k,3}}\ .
	\end{dcases}
\end{equation}
The LLR for bit $b_{k,i}$ of user $(k,i)$, $i = A, B$ is 
\begin{equation}
\begin{aligned}
e_{LLR}(b_{k,i}) = \sum_{l=0}^{L-1} \alpha_{l}  e^{l}_{LLR}(b_{k,i}),
\label{equ:CDMAMLDTcomb}
\end{aligned}
\end{equation}
where
\begin{equation}
\begin{aligned}
\alpha_{l} = \dfrac{|h_{k,p,l}|^{2}} {\sum_{l=0}^{L-1} |h_{k,p,l}|^{2}}.
\end{aligned}
\end{equation}

\subsection{Systems with Hadamard Walsh codes}
Hadamard Walsh (HW) codes can be used in the CDMA applications employing orthogonal signature sequences.  We set $K$ = $J$.  With these codes, the interference from other users can be removed perfectly for $L=1$. Hence, in case of single-path Rayleigh fading, the BER performances for CDMA using HW codes with $P$ = $2$ and $3$ are exactly the same as those derived in Section II.B and shown in Fig. \ref{fig:analysis_uncoded}. That means that we can double or triple the number of users in the CDMA system using HW codes with only slight BER degradation.

For $L > 1$, the term $w_{mp}$ is significant since for the HW code, the autocorrelation of a sequence with its shift may be significant.  Hence, the BER performances will be very poor.



\subsection{Systems with m-sequences}
\begin{figure}[H]
	\centering
	\includegraphics[width=17cm, height=12cm]{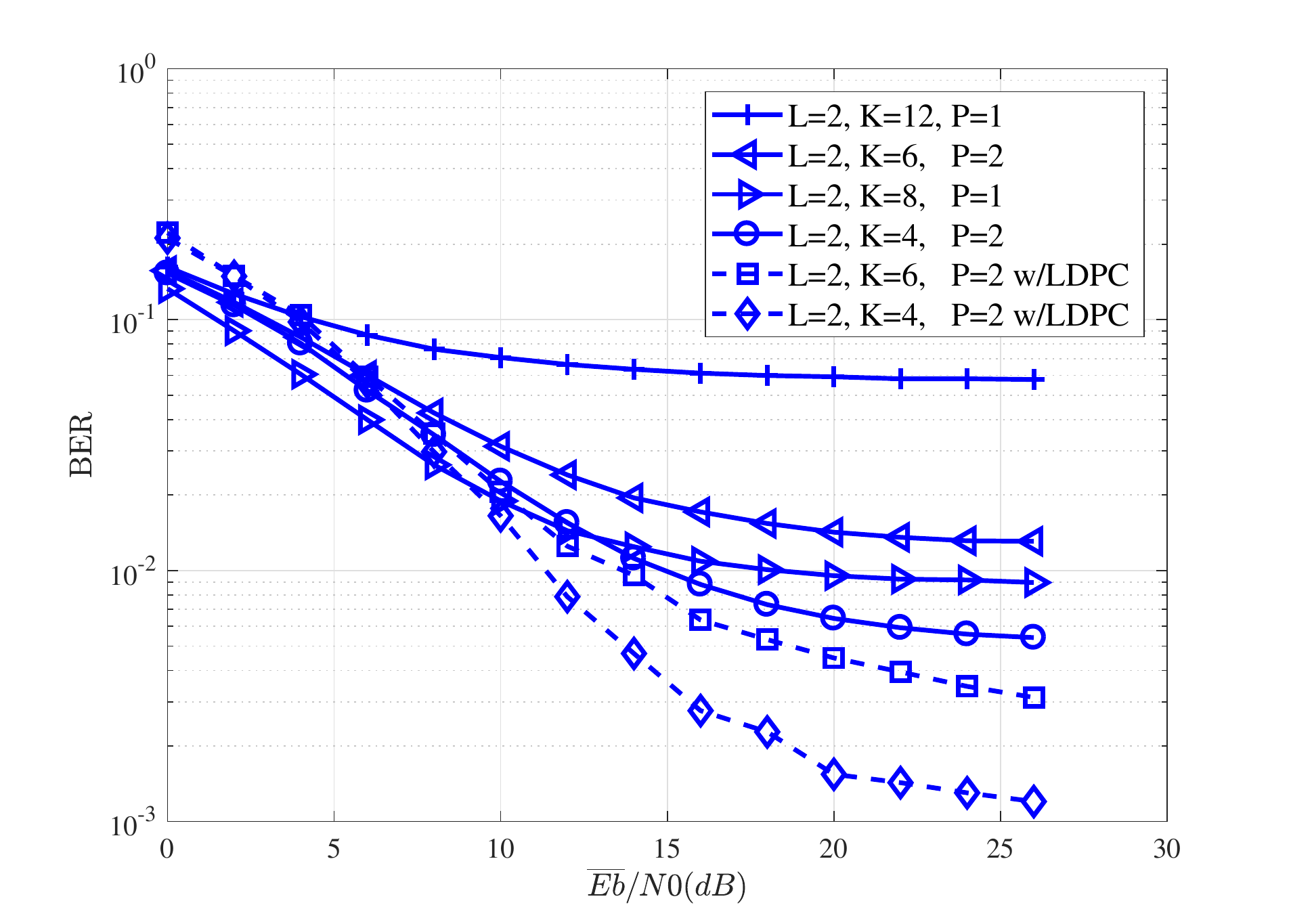}
	\caption{CDMA system with m-sequences of length $J$ = 15 over a 2-path Rayleigh fading channel. MLDT receiver for $P$ = 2.}
	\label{fig:sim_LDPC_CDMA_m}
\end{figure}

We replace HW codes by m-sequences (maximum length sequences) in multipath channels, where the autocorrelation of a sequence with its shift is small.
 In Fig. \ref{fig:sim_LDPC_CDMA_m}, the BER performances of the m-sequence CDMA system, in the two-path Rayleigh environment with $J$ = 15, $P$ = 1 and 2 respectively are compared, where the two paths have equal average powers.  We see that using the MLDT receivers is not able to perform well in the non-orthogonal environment.  The multipath interference severely affect the BER performances. Note that if $KL > J$ which implies that the number of m-sequences in use multiplied by the number of paths exceeds  the number of available m-sequences, then the BER performances will be extremely poor.  To tackle the multipath interference, we consider using the (1008, 504) $(3,6)$-regular LDPC code \cite{Mackay} and binary SPA decoders following the MLDT at the receiver.  
 In Fig. \ref{fig:sim_LDPC_CDMA_m}, we can see that the BER performances for the 2-path CDMA system can be somewhat improved by using the LDPC code.  However, the BER performances are not satisfactory even for only $P$ = 2.  We will resort to  multi-carrier CDMA in the following subsection.

\subsection{Multi-Carrier Direct-Sequence CDMA (MC-DS-CDMA) System} \label{sec:frequency_domain}
To maintain the orthogonality of HW code over multipath channels, the MC-DS-CDMA system is considered. 

\subsubsection{System Model}
\begin{figure}[H]
		\centering
		\includegraphics[width=1\columnwidth]{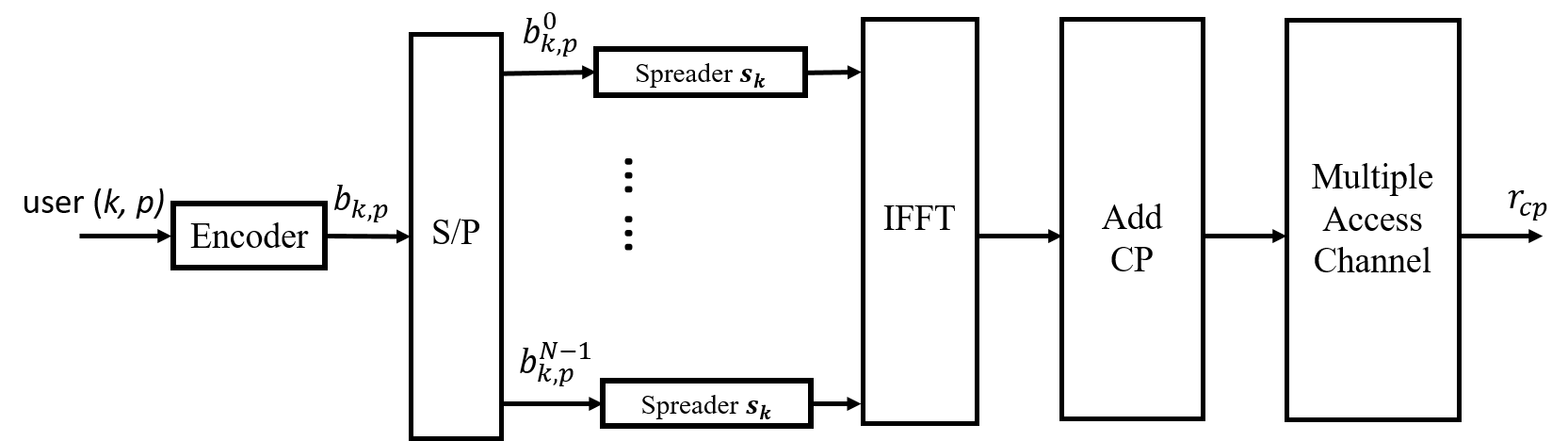}
		\caption{The transmitter of the MC-DS-CDMA system for user $(k,p)$.}
		\label{fig:4-1}
\end{figure}

The transmitter structure of the MC-DS-CDMA system is illustrated in Fig. \ref{fig:4-1}. The input sequence of user $(k,p)$ is ${\bf b}_{k,p}$ = $[b^{0}_{k,p}, ... , b^{n}_{k,p}, ..., b^{N-1}_{k,p}]$.  For $p = 1, ..., P$, the bit $b^{n}_{k,p}$  of user $(k,p)$ is spread by the signature sequence  ${\bf s}_k$ = $[s_{k}(1), ..., s_{k}(J)]$ for the $n$th subcarrier.  We have ${\bf X^{n}_{k,p}}$ = $[X^{n}_{k,p}(1), ..., X^{n}_{k,p}(J)]$, where $X^{n}_{k,p}(j) = b^{n}_{k,p}s_{k}(j)$, which is then processed by an $N$-point IFFT (inverse fast Fourier transform). At the end, the cyclic prefix (CP) is added to cope with the multipath effect.

\begin{figure}[H]
		\centering
		\includegraphics[width=16cm, height=3.3cm]{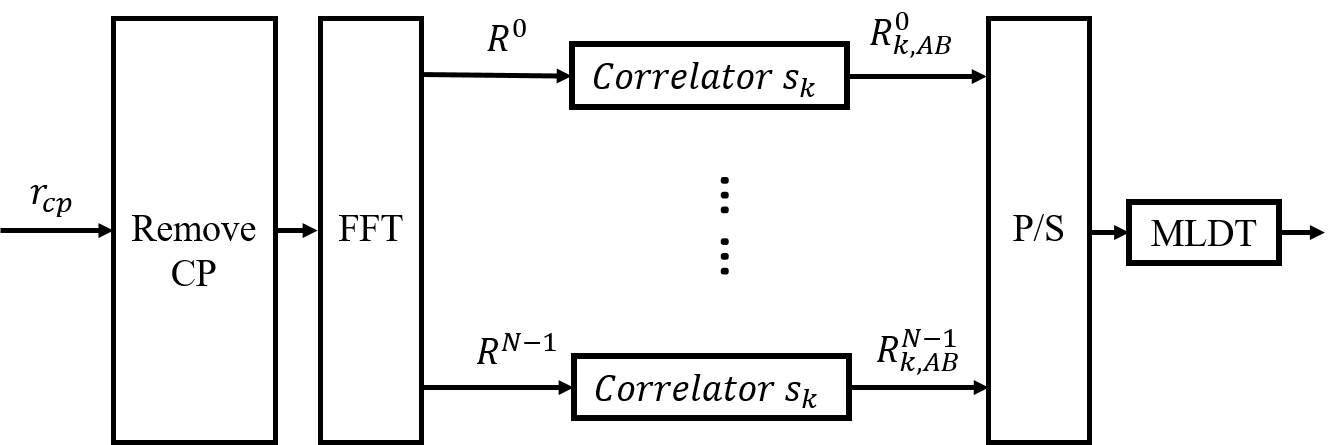}
		\caption{The receiver of the MC-DS-CDMA system for user $(k,p)$.}
		\label{fig:4-2}
\end{figure}

Fig. \ref{fig:4-2} shows the receiver structure of user $(k,p)$. The received signal in the frequency domain after removing CP can be expressed as ${\bf R}^{n}$ = $[R^{n}(1), ..., R^{n}(J)]$, 
\begin{equation}\label{equ:MCDSCDMA_r}
R^{n}(j)=\sum_{k=1}^K \sum_{p=1}^PH^{n}_{k,p}X^{n}_{k,p}(j) + W(j),\ 1\leq j\leq J,
\end{equation}
where $H^{n}_{k,p}$ is the channel coefficient of user $(k,p)$ on the $n$-th subcarrier and $W(j)$ is the Gaussian noise. For $P$ = 2, after de-spreading, we have
\begin{equation}
{R}^{n}_{k,AB}=\frac{1}{J} \sum_{j=1}^J R^{n}(j)s_{k}(j).
\end{equation}

For $P$ = 2, the corresponding table for the superimposed transmitted signal of the collided user $A$ and user $B$ in the frequency domain are slightly modified and are shown in TABLE \ref{table:superimposed txx}. For $P$ = 3, $S_{ABC}(i)$, $i = 0, 1, 2, ..., 7$ can be similarly obtained.
\begin{table}[H]
\caption{The superimposed signal and the corresponding data in the frequency domain.}
	\begin{center}
		\begin{tabular}{ | >{\centering\arraybackslash}m{\columnwidth/30}  | >{\centering\arraybackslash}m{\columnwidth/10}  | >{\centering\arraybackslash}m{\columnwidth/10} | >{\centering\arraybackslash}m{\columnwidth/10} | >{\centering\arraybackslash}m{\columnwidth/10} | >{\centering\arraybackslash}m{\columnwidth/6} | }
		\hline
 		$i$ & $b^n_{k,A}$ & $b^n_{k,B}$ & $X^n_{k,A}$ & $X^n_{k,B}$ & $S_{k,AB}(i)$ \\ \hline
 		0 & 0 & 0 & 1 & 1 & $H^n_{k,A}+H^n_{k,B}$ \\ \hline 
 		1 & 0 & 1 & 1 & -1 & $ H^n_{k,A}-H^n_{k,B}$ \\ \hline 
 		2 & 1 & 0 & -1 & 1 & $-H^n_{k,A}+H^n_{k,B}$ \\ \hline 
 		3 & 1 & 1 & -1 & -1 & $-H^n_{k,A}-H^n_{k,B}$ \\ \hline
		\end{tabular}
		\label{table:superimposed txx}
	\end{center}
\end{table}

\subsubsection{Simulation Results}
\begin{figure}[H]
	\centering
	\includegraphics[width=16cm, height=11cm]{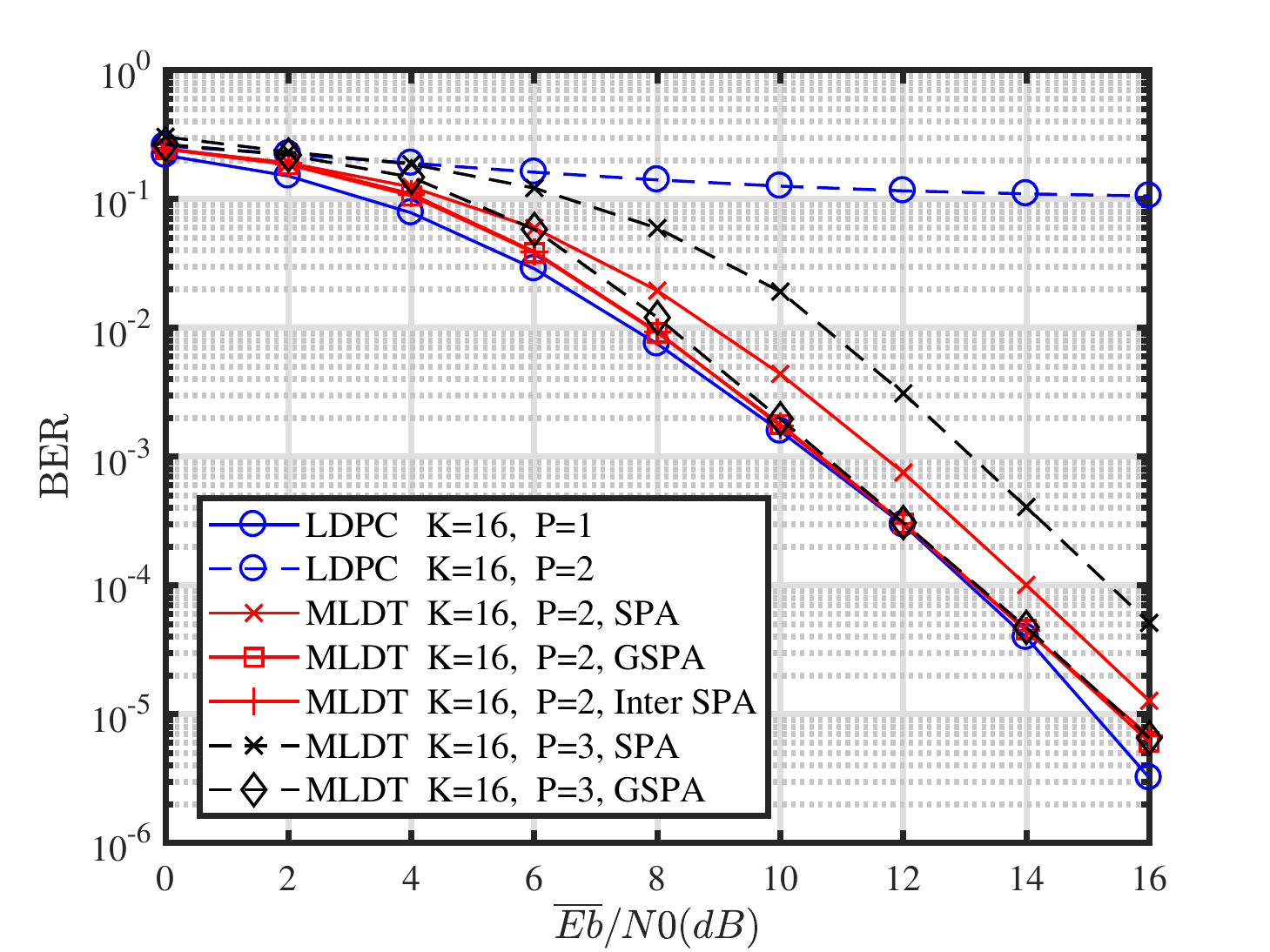}
	\caption{$(1008, 504)$ LDPC-coded MC-DS-CDMA system in multipath Rayleigh fading Channels.}
	\label{fig:sim_LDPC_MCDSCDMA_1008}
\end{figure}

In the simulation, we consider an LDPC-coded MC-DS-CDMA system in multipath Rayleigh fading channels of which the path length is $L=5$ and each path has the same average power.  We set $K$ = $J$ = 16. The $(1008,504)$ LDPC code is used.  The FFT size and the CP length are set to 16 and 4, respectively.  The simulated BER performances are provided
in Fig. \ref{fig:sim_LDPC_MCDSCDMA_1008}. 

It is interesting to see that using MLDT with a single GSPA decoder for either $P$ = 2 or $P$ = 3 can obtain BER performances very close to those obtained for only $P$ = 1.  Hence, the number of users can be doubled from 16 to 32 or tripled from 16 to 48. This is a significant advantage.

\begin{figure}[H]
	\centering
	\includegraphics[width=16cm, height=2.5cm]{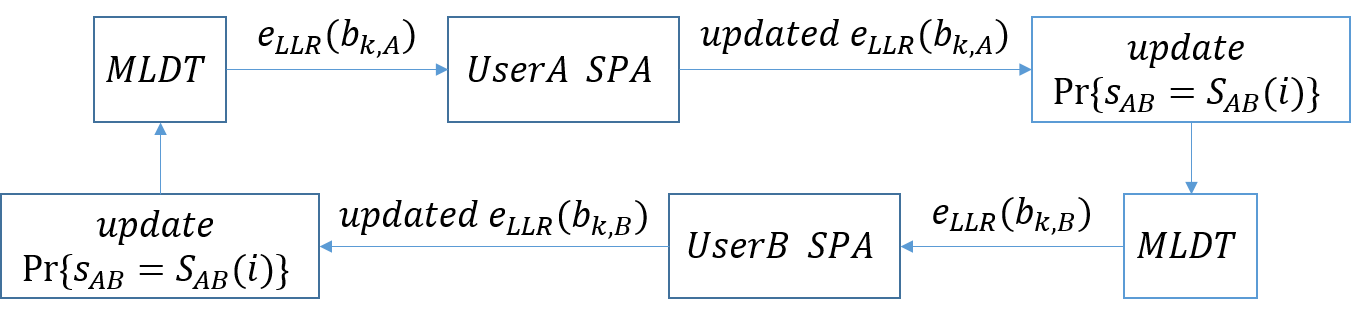}%
	\caption{Illustration of MLDT with inter SPA.}
	\label{fig:interSPA_MLDT}
\end{figure}

Compared to MLDT with a single GSPA decoder, we see that MLDT with multiple SPA decoders can obtain somewhat inferior BER performances for both $P$ = 2 and $P$ = 3.  However, using multiple SPA decoders has the advantage of lower decoding complexity.  The BER performances of MLDT with multiple SPA decoders for $P$ = 2 can be improved by employing the interchange of LLR values  between the two SPA decoders. 
The modified  MLDT receiver followed by two SPA decoders, which is denoted as MLDT with inter SPA, is depicted Fig. \ref{fig:interSPA_MLDT}.  The output of the MLDT receiver is first processed by the SPA decoder of user $A$, which generates  updated $e_{LLR}({b}_{k,A})$ to update $\text{Pr}\{s_{AB} = S_{AB}(i)\}$.  Then, the MLDT is able to generate $e_{LLR}({b}_{k,B})$ by (\ref{equ:recalculate_p}), which is then processed by the SPA decoder of user $B$. We can repeat these steps to obtain improved LLR values.  From Fig. \ref{fig:sim_LDPC_MCDSCDMA_1008}, we see that for $P$ = 2, the BER performances of MLDT with inter SPA  are very close to those of MLDT with a single GSPA.

\begin{equation}\label{equ:recalculate_p}
\begin{dcases}
&p_0 = \frac{\beta}{2\pi\sigma^2}\exp(-\frac{|r_{AB}-S_{AB}(0)|^2}{2\sigma^2})(1-\frac{1}{1+\exp[e_{LLR}(b_{k,A})]})\\
&p_1 = \frac{\beta}{2\pi\sigma^2}\exp(-\frac{|r_{AB}-S_{AB}(1)|^2}{2\sigma^2})(1-\frac{1}{1+\exp[e_{LLR}(b_{k,A})]})\\
&p_2 = \frac{\beta}{2\pi\sigma^2}\exp(-\frac{|r_{AB}-S_{AB}(2)|^2}{2\sigma^2})(\frac{1}{1+\exp[e_{LLR}(b_{k,A})]})\\
&p_3 = \frac{\beta}{2\pi\sigma^2}\exp(-\frac{|r_{AB}-S_{AB}(3)|^2}{2\sigma^2})(\frac{1}{1+\exp[e_{LLR}(b_{k,A})]})\\
&e_{LLR}({b}_{k,B}) = \ln\frac{p_0+p_2}{p_1+p_3},
\end{dcases}
\end{equation}
where $\beta$ is a normalization factor.



\section{Concluding Remarks} \label{sec:con}
We propose to use an MLDT technique to allow multiple users which transmit signals over independent fading channels to share the same resource.  Both BER analysis and simulation for the uncoded system show that using MLDT can double or triple the number of users for multiple access with the price of some degradation of BER performances.  Capacity analysis is provided to show the possible gains that we can achieve.  Designs of Raptor coded systems using MLDT over the single-path quasi-static Rayleigh fading channels and LDPC-coded multi-carrier direct-sequence CDMA systems using MLDT over the multi-path quasi-static Rayleigh fading channels show that both appreciable coding gains and increased number of users can be obtained.


%

\ifCLASSOPTIONcaptionsoff
  \newpage
\fi

\bibliographystyle{IEEEtran}
\bibliography{ref}

\begin{thebibliography}{10}
\providecommand{\url}[1]{#1}
\csname url@samestyle\endcsname
\providecommand{\newblock}{\relax}
\providecommand{\bibinfo}[2]{#2}
\providecommand{\BIBentrySTDinterwordspacing}{\spaceskip=0pt\relax}
\providecommand{\BIBentryALTinterwordstretchfactor}{4}
\providecommand{\BIBentryALTinterwordspacing}{\spaceskip=\fontdimen2\font plus
\BIBentryALTinterwordstretchfactor\fontdimen3\font minus
  \fontdimen4\font\relax}
\providecommand{\BIBforeignlanguage}[2]{{%
\expandafter\ifx\csname l@#1\endcsname\relax
\typeout{** WARNING: IEEEtran.bst: No hyphenation pattern has been}%
\typeout{** loaded for the language `#1'. Using the pattern for}%
\typeout{** the default language instead.}%
\else
\language=\csname l@#1\endcsname
\fi
#2}}
\providecommand{\BIBdecl}{\relax}
\BIBdecl

\bibitem{Li06}
L.~Ping, L.~Liu, K.~Wu, and W.~K. Leung, ``Interleave division
  multiple-access,'' \emph{IEEE Transactions on Wireless Communications},
  vol.~5, no.~4, pp. 938--947, April 2006.

\bibitem{Saito13}
Y.~Saito, Y.~Kishiyama, A.~Benjebbour, T.~Nakamura, A.~Li, and K.~Higuchi,
  ``{Non-orthogonal multiple access (NOMA) for cellular future radio access},''
  in \emph{IEEE VTC}, June 2013, pp. 1--5.

\bibitem{Hoshyar08}
R.~Hoshyar, F.~P. Wathan, and R.~Tafazolli, ``{Novel low-density signature for
  synchronous CDMA systems over AWGN channel},'' \emph{IEEE Transactions on
  Signal Processing}, vol.~56, no.~4, pp. 1616--1626, Apr. 2008.

\bibitem{Nikopour13}
H.~Nikopour and H.~Baligh, ``Sparse code multiple access,'' in \emph{2013 IEEE
  24th Annual International Symposium on Personal, Indoor, and Mobile Radio
  Communications (PIMRC)}, Sept 2013, pp. 332--336.

\bibitem{LDPC}
M.~C. Davey and D.~J.~C. MacKay, ``{Low density parity check codes over
  GF(q)},'' in \emph{Information Theory Workshop}, June 1998, pp. 70--71.

\bibitem{Kschischang01}
F.~R. Kschischang, B.~J. Frey, and H.~A. Loeliger, ``Factor graphs and the
  sum-product algorithm,'' \emph{IEEE Transactions on Information Theory},
  vol.~47, no.~2, pp. 498--519, Feb. 2001.

\bibitem{PLNC}
D.~Wubben and Y.~Lang, ``Generalized sum-product algorithm for joint channel
  decoding and physical-layer network coding in two-way relay systems,'' in
  \emph{IEEE Globecom Proceedings, Miami, Florida, USA}, Dec. 2010, pp. 1--5.

\bibitem{Shokrollahi}
A.~Shokrollahi, ``Raptor codes,'' \emph{IEEE Transactions on Information
  Theory}, vol.~52, no.~6, pp. 2551--2567, June 2006.

\bibitem{Goldsmith05}
A.~Goldsmith, \emph{Wireless communications}.\hskip 1em plus 0.5em minus
  0.4em\relax Cambridge University Press, 2005.

\bibitem{Mackay}
\BIBentryALTinterwordspacing
{David J.C. MacKay}. {Encyclopedia of Sparse Graph Codes}. [Online]. Available:
  \url{http://www.inference.org.uk/mackay/codes/data.html}
\BIBentrySTDinterwordspacing

\bibitem{Kuo}
S.~H. Kuo, Y.~L. Guan, S.~K. Lee, and M.~C. Lin, ``A design of physical-layer
  raptor codes for wide snr ranges,'' \emph{IEEE Communications Letters},
  vol.~18, no.~3, pp. 491--494, March 2014.

\bibitem{Jayasooriya17}
S.~Jayasooriya, M.~Shirvanimoghaddam, L.~Ong, and S.~J. Johnson, ``Analysis and
  design of raptor codes using a multi-edge framework,'' \emph{IEEE
  Transactions on Communications}, vol.~65, no.~12, pp. 5123--5136, Dec 2017.

\end{thebibliography}

\end{document}